\documentclass[
aps,
twocolumn,
prl,
groupedaddress,
superscriptaddress,
letterpaper,
amsfonts,
floatfix
]
{revtex4-1}
\RequirePackage{amsmath,amssymb,bm,wasysym}
\RequirePackage{graphicx}
\RequirePackage{hyperref}
\RequirePackage{braket}
\hypersetup{
breaklinks={true},
citecolor={blue},
colorlinks={true},
linkcolor={blue},
pdfcreator={\LaTeX and\flqq hyperref\frqq},
pdffitwindow={true},
pdfmenubar={true},
pdfpagelayout={OneColumn},
pdfstartview={FitH},
pdftoolbar={true},
plainpages={false},
pdfpagemode={UseThumbs},
}
\DeclareMathAlphabet{\mathpzc}{OT1}{pzc}{m}{it}
\renewcommand{\vec}[1]{\boldsymbol{#1}}

\newcommand{\bea}{\begin{eqnarray*}}
\newcommand{\eea}{\end{eqnarray*}}
\newcommand{\bne}{\begin{equation*}}
\newcommand{\ede}{\end{equation*}}

\newcommand{\bnen}{\begin{equation}}
\newcommand{\eden}{\end{equation}}
\newcommand{\bean}{\begin{eqnarray}}
\newcommand{\eean}{\end{eqnarray}}
\newcommand{\bsen}{\begin{subequations}}
\newcommand{\esen}{\end{subequations}}

\newcommand{\bna}{\begin{array}}
\newcommand{\eda}{\end{array}}
\newcommand{\bnm}{\begin{enumerate}}
\newcommand{\edm}{\end{enumerate}}
\newcommand{\bni}{\begin{itemize}}
\newcommand{\edi}{\end{itemize}}

\renewcommand{\vec}[1]{\text{\boldmath{$ #1 $}}}

\begin{document}
\pagestyle{plain}
\title{Hyperfine-assisted fast electric control of
dopant nuclear spins in semiconductors}
\author{P\'eter Boross}
\affiliation{Institute of Physics, E\"{o}tv\"{o}s University, 1518 Budapest, Hungary}
\author{G\'abor Sz\'echenyi}
\affiliation{Institute of Physics, E\"{o}tv\"{o}s University, 1518 Budapest, Hungary}
\author{Andr\'{a}s P\'{a}lyi}
\thanks{Corresponding author: palyi@mail.bme.hu}
\affiliation{Department of Physics, Budapest University of Technology and Economics, 1111 Budapest, Hungary}
\affiliation{MTA-BME Condensed Matter Research Group, 
Budapest University of Technology and Economics, 1111 Budapest, Hungary}
\pacs{
}
\begin{abstract}
Nuclear spins of dopant atoms in semiconductors are 
promising candidates as quantum bits, 
due to the long lifetime of their quantum states. 
Conventionally, coherent control of nuclear spins
is done using ac magnetic fields. 
Using the example of
a phosphorus atom in silicon, 
we theoretically 
demonstrate that hyperfine interaction can enhance the speed
of magnetic control: the electron on the donor amplifies  the ac
magnetic field felt by the 
nuclear spin. 
Based on that result, we show that hyperfine interaction also
provides a means to control the nuclear spin efficiently using 
an ac electric field, in the presence of intrinsic or artificial 
spin-orbit interaction.
This electric control scheme
is especially efficient and noise-resilient in a hybrid dot-donor
system holding two electrons 
in the presence of an inhomogeneous magnetic field. 
The mechanisms proposed here could be used as building blocks
in  nuclear-spin-based electronic quantum information 
architectures. 
\end{abstract}
\maketitle

\emph{Introduction.}
The nuclear spin of a phosphorus  (P) atom in silicon (Si)
is a highly coherent two-level system\cite{Muhonen,Saeedi},
and hence is
a competitive candidate for representing a qubit
in quantum information 
processing\cite{Kane,Zwanenburg}.
Resonant control of  a single nuclear-spin qubit 
have been demonstrated using ac magnetic fields in the spirit of 
nuclear magnetic resonance\cite{Pla_nuc};
initialization and readout can be performed using the 
donor electron spin\cite{Morello,Pla_electron}. 

Similarly to the case of electron spin qubits, for nuclear spins it
would also be beneficial to substitute the ac magnetic  control 
with ac electric  control: 
this could lead to simplified sample design, 
spatially confined control fields, 
lower power requirements, 
higher qubit densities,
shorter gate times, and
an opportunity to couple nuclear spins electrically to 
each other or to 
electromagnetic resonators\cite{Tosi-nuclearspin,Sigillito}.
One challenge to achieve these for the P:Si system is that
P has a spin-$1/2$ nuclear spin, which does not couple
directly to the electric field\cite{Slichter}. 
Nevertheless, recent activities advance toward this goal 
by proposing a combined electric-magnetic control scheme 
in a dot-donor system\cite{Tosi-nuclearspin}, 
by demonstrating electric control of the hyperfine 
interaction\cite{Laucht}
and nuclear spin states 
in an ensemble of nuclear spins in Si\cite{Sigillito} and
in a single-molecule magnet\cite{Thiele,Godfrin}. 

In this work, we propose and analyze a mechanism 
to control the P nuclear spin 
efficiently using ac electric fields. 
The mechanism relies on hyperfine interaction:
we utilize the donor-bound electron as a
quantum transducer between the electric field and the nuclear spin.
Actually, this transducer effect can be exploited to speed up
qubit control already
in the case of magnetic drive, which is a conceptually simpler
scenario; therefore, first we discuss how  it 
works in the case of magnetic drive. 

\begin{figure}
\centering
\includegraphics[width=0.9\columnwidth]{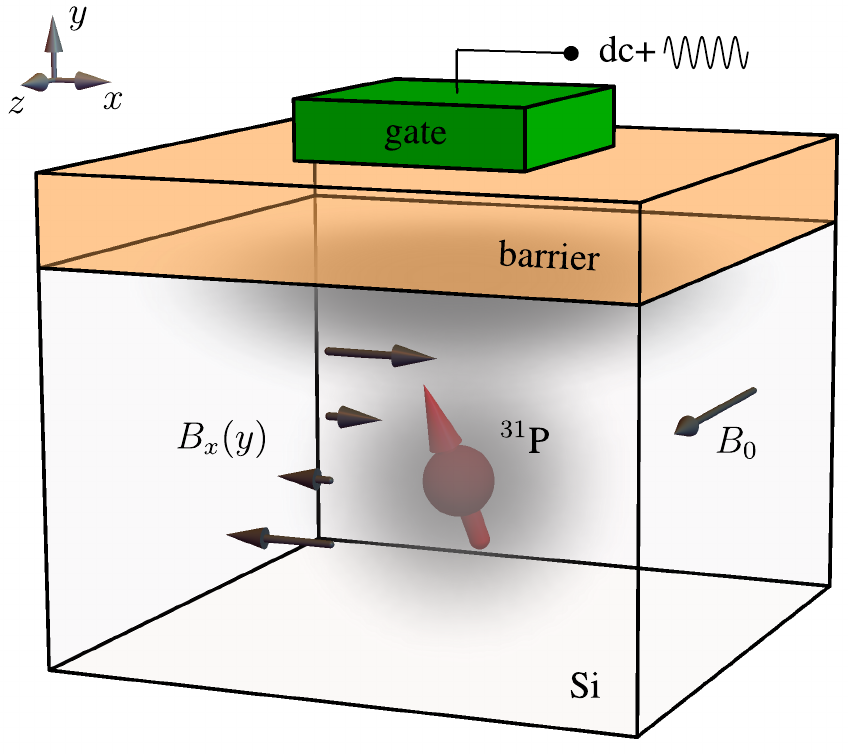}
\caption{
\textbf{
Electrically driven nuclear spin resonance of a phosphorus atom
in a silicon dot-donor system.}
The dc gate voltage is used to balance the donor's electron
on a bonding orbital (gray cloud) of the artificial molecule 
formed by the dot-donor system.
The ac gate-voltage component oscillates 
the electron vertically.
Due to these charge oscillations and the presence of the
inhomogeneous magnetic field $B_x(y)$, 
the electron spin  on the donor acquires an
oscillating $x$ component, which drives the nuclear spin (red arrow) 
via
the hyperfine interaction.
\label{fig:setup}}
\end{figure}

\emph{Hyperfine-assisted magnetic control of the P:Si nuclear spin.}
First, consider a single ionized P atom in Si, in a 
homogeneous static magnetic field $B_0$ along the $z$ axis,
driven resonantly by an ac magnetic field along the $x$ axis, with
amplitude $B_\text{ac}$ and frequency $f$.
Dynamics is described by the Hamiltonian 
$H_{B,\text{n}} + H_\text{d,n}(t)$, where
$H_{B,\text{n}} = - h \gamma_\text{n} B_0 I_z$
and 
$H_\text{d,n}(t) = - h \gamma_\text{n} B_\text{ac} \sin (2\pi f t) I_x$,
$\gamma_\text{n} = 17.23 \, \text{MHz}/\text{T} $ 
is the nuclear gyromagnetic ratio, 
and $\vec I = (I_x,I_y,I_z)$ is the spin-$1/2$ nuclear spin operator.
When driven resonantly ($f=f_L^{(i)} \equiv \gamma_\text{n} B_0$),
the nuclear spin performs complete Rabi oscillations with  
Rabi frequency 
$f^{(i)}_\text{R} = \frac{1}{2} \gamma_\text{n} B_\text{ac}$.

Consider how this result changes when the donor is not ionized, 
but neutral, i.e., it has a single electron (1e) occupying the ground-state
donor orbital.
The Hamiltonian is
\bean
H(t) = H_{B,\text{e}} + H_{B,\text{n}} + H_\text{hf}
+H_\text{d,e}(t) + H_\text{d,n}(t).
\eean
Here, 
$H_{B,\text{e}} =
h \gamma_\text{e} B_0 S_z$,
$H_\text{hf} = A \vec S \cdot \vec I$, and
$H_\text{d,e} =
h \gamma_\text{e} B_\text{ac} \sin(2\pi f t) S_x $,
whereas $\gamma_\text{e} = 27.97\, \text{GHz}/\text{T}$ 
is the electron 
gyromagnetic ratio, $A/h= 117 \, \text{MHz}$ is the hyperfine 
coupling strength,
and $\vec S = (S_x,S_y,S_z)$ is the electron spin operator. 
Consider the experimentally relevant case\cite{Pla_nuc} when
the electronic Zeeman splitting dominates the hyperfine 
strength, $h \gamma_\text{e} B_0 \gg A$.  
Assume that the system is in its ground state, well approximated by 
the state 
$\ket{\downarrow \Uparrow}$ with the 
electron spin pointing down and the nuclear spin pointing up,
before the driving starts.
Then, the nuclear Larmor frequency is 
$f_L^{(\downarrow)}
\approx \frac{A}{2h} + \gamma_\text{n} B_0 \ll \gamma_\text{e} B_0 $.
Upon driving at the nuclear Larmor frequency,
due to $f = f_L^{(\downarrow)} \ll \gamma_\text{e} B_0$, the electron 
spin will adiabatically follow the direction of the instantaneous
magnetic field: 
$\braket{\vec S}_t\approx - \frac 1 2 
\left( \frac{B_\text{ac}}{B_0}  \sin (2\pi f t), 0, 1\right)$. 
This electron spin dynamics will in turn create an 
additional transverse driving field (Knight field) on the 
nuclear spin via the hyperfine interaction: 
$H'_\text{d,n}(t) = A \braket{S_x}_t I_x = - \frac {A B_\text{ac}}{2 B_0} \sin(2\pi f t) I_x$.
As a consequence of this extra driving term, 
the Rabi frequency is increased with 
respect to the ionized case:
$f_\text{R}^{(\downarrow)} = \frac{1}{2}
\left(
\gamma_\text{n} + \frac{A}{2hB_0}
\right) B_\text{ac}$.
A similar consideration shows that for an up-spin electron, the
Rabi frequency is 
$f_\text{R}^{(\uparrow)} = \frac{1}{2}
\left(
\gamma_\text{n} - \frac{A}{2hB_0}
\right) B_\text{ac}$.

An analogous effect in nitrogen-vacancy 
defects in diamond was termed 
`hyperfine-enhanced nuclear gyromagnetic
ratio', and was characterized theoretically and 
experimentally\cite{Childress,Smeltzer,Sangtawesin14,Chen,Sangtawesin16}.
For the P:Si system, a straightforward experimental proof of 
this hyperfine-assisted nuclear spin control 
could be obtained by extending the analysis presented
in Fig.~3 of \cite{Pla_nuc}. 
E.g., using the nuclear-spin Rabi
oscillations of the ionized P, the ac magnetic field amplitude at
the donor position can be determined, which can be used to predict
the nuclear-spin Rabi frequency for the neutral P with
an up-spin electron. 

\emph{Hyperfine-assisted electric control of the P:Si nuclear spin.}
The previous mechanism suggests the possibility of ac electric control, 
in the case when the ac
 electric field can modulate the local instantaneous Knight field
 at the position of the P nucleus. 
We demonstrate that this
 modulation can be achieved in a
 two-site system controlled by gate electrodes, e.g., 
 in a dot-donor system\cite{Lansbergen,Tosi_natcomm,Urdampilleta,HarveyCollard,Boross,HarveyCollard2,Rudolph}, in the presence of an inhomogeneous
 magnetic field\footnote{Similar setups have been studied for the purpose of 
electric control of electron spins\cite{Tokura,PioroLadriere}}.

\begin{figure*}
\centering
\includegraphics[width=2.0\columnwidth]{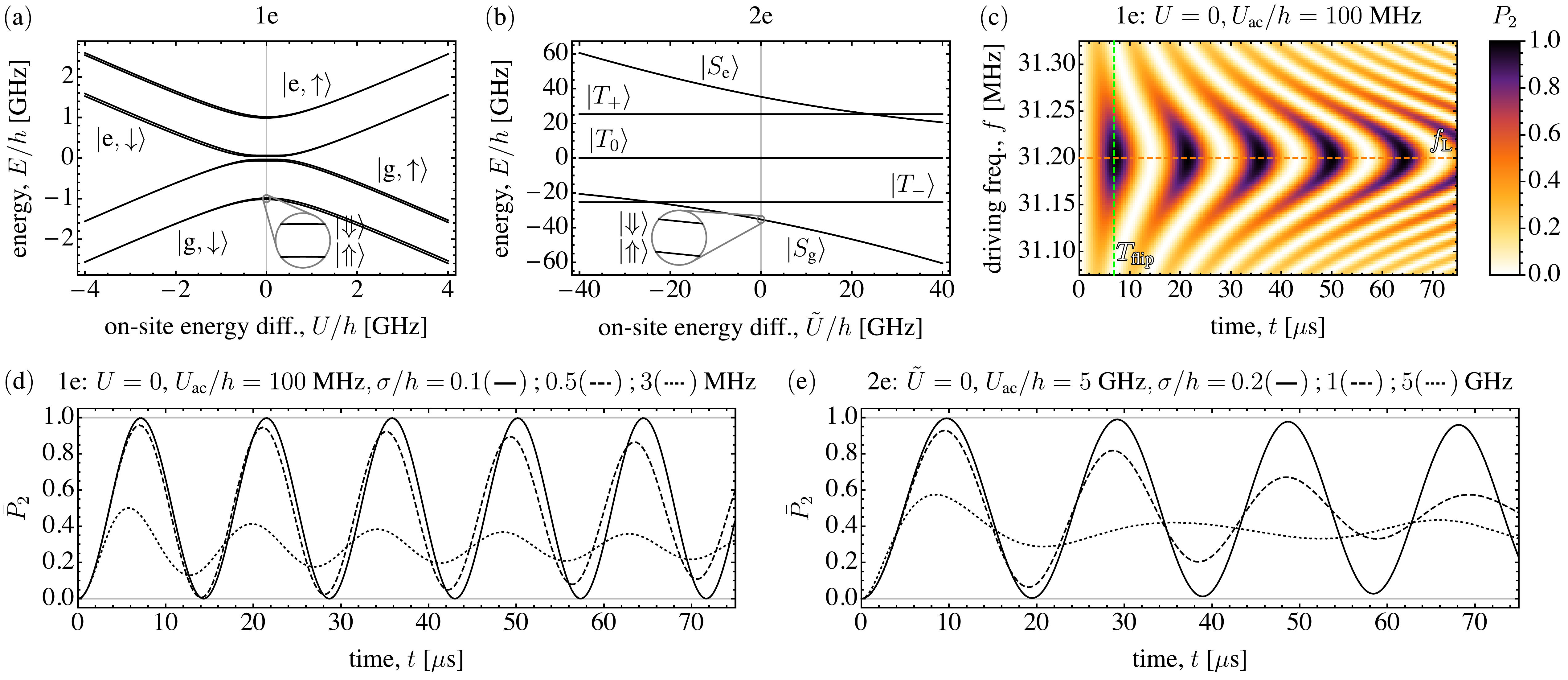}
\caption{
{\bf Electrically driven nuclear-spin Rabi oscillations and 
their damping
due to charge noise.}
(a,b) shows the energy level structure of
the single-electron (1e) and two-electron (2e) setups, 
as functions of energy detuning from the (1,0)-(0,1) 
and (1,1)-(0,2) tipping points, respectively. 
(c) Nuclear-spin Rabi oscillations induced
by electrically driving a single electron at the tipping point $U=0$.
Time-dependent occupation probability $P_2(t,f)$ of the excited 
state of the nuclear-spin qubit
shows the standard Chevron pattern of magnetic 
resonance, in the absence of noise. 
Green dashed line shows the spin-flip time 
$T_\text{flip} = 1/(2f_\text{R})$ 
predicted by Eq.~\eqref{eq:rabi}.
Orange dashed line shows the numerically calculated 
nuclear-spin Larmor frequency $f_\text{L}$, i.e., the  
seperation between the two lowest eigenvalues
of the static Hamiltonian. 
(d,e) Damping of the  nuclear-spin Rabi oscillations
due to different strengths of charge noise ($\sigma$),
in the 1e and 2e setups, 
for resonant drive.
$\bar{P}_2(t)$ is the noise-averaged occupation probability
of the excited state of the nuclear-spin qubit, see Eq.~\eqref{eq:noiseaverage}. 
The 2e setup is three orders of magnitude more resilient 
to charge noise than the 1e setup. 
Parameters:
$\beta = 0.47\, \text{mT}/\text{nm}$, 
$d = 15\, \text{nm}$.
(a,c,d)
$V_\text{t}/h = 1\, \text{GHz}$,
$B_0 =35.7\, \text{mT}$ 
(b,e)
$V_\text{t}/h = 50\, \text{GHz}$,
$B_0 = 906.5\, \text{mT}$.
\label{fig:numericalresults}}
\end{figure*}

First, we consider the case when a single electron 
is confined in the dot-donor system (Fig.~\ref{fig:setup}).
The orbital degree of freedom is described
in the two-dimensional Hilbert space spanned by the ground-state
orbital localized in the dot ($\ket{\text{i}}$) 
and that localized on the donor ($\ket{\text{d}}$).
The distance between the centers of these orbitals is $d$.
The magnetic-field inhomogeneity is
characterized by the field gradient $\beta$, and  
the magnetic field takes the values
$\left(\pm \frac{\beta d}{2},0,B_0\right)$ 
on the dot and the donor, respectively.
Then, the 
minimal model of this setup\cite{Tosi_natcomm,Boross,Tosi-nuclearspin} can be written as
 \bean
 \label{eq:model}
 H = H_\text{o} + H_{B,\text{e}} + H_{B,\text{n}}
 + H_\text{hf} + H_{\mu,\text{e}} + H_{\mu,\text{n}}
 + H_d(t), \nonumber \\
 \eean
with the orbital Hamiltonian  
$H_\text{o} = \frac U 2 \sigma_z + \frac{V_\text{t}}{2} \sigma_x$,
the hyperfine interaction 
$H_\text{hf} =A n_\text{d} \vec S \cdot \vec I$,
the inhomogeneous magnetic field
\begin{subequations}
\bean
\label{eq:einhom}
H_{\mu,\text{e}} &=& h \gamma_\text{e} \frac{\beta d}{2} \sigma_z
S_x
\\
H_{\mu,\text{n}} &=& h \gamma_\text{n} \frac{\beta d}{2} I_x,
\eean
\end{subequations}
and the electric drive
$H_\text{d}(t) = \frac{U_\text{ac}}{2}\sigma_z \sin (2\pi f t)$.
Here, $U$ is the gate-tunable on-site energy difference between
$\ket{\text{i}}$ and $\ket{\text{d}}$, $V_\text{t}$ is the tunnel coupling
between them, $\sigma_{x,y,z}$ are the Pauli matrices acting
on the orbital degree of freedom (e.g., $\sigma_z = \ket{\text{i}} \bra{\text{i}}- \ket{\text{d}}\bra{\text{d}}$), 
$n_\text{d} \equiv (1-\sigma_z)/2$ is the electron number on
the donor, 
and $U_\text{ac} = e E_\text{ac} d$ is the on-site energy difference 
induced by 
an ac electric field $E_\text{ac}$ along the dot-donor axis, 
created via an ac voltage excitation of the
gate electrode. 
The 8 energy eigenvalues of this Hamiltonian are shown in 
Fig.~\ref{fig:numericalresults}a as a function of the
on-site energy difference $U$
(see caption for parameters); 
the lowest two branches labelled $\ket{\Downarrow}$
and $\ket{\Uparrow}$
correspond to the basis states of the nuclear-spin qubit.

To illustrate that electric driving $H_\text{d}(t)$ results in a
time-dependent Knight field for the donor nuclear spin, and 
thereby induces nuclear-spin Rabi oscillations, we 
consider a simple case: 
when the electron is balanced 
at the tipping point between the $\ket{\text{i}}$ 
and $\ket{\text{d}}$ orbitals ($U=0$, gray vertical line in 
Fig.~\ref{fig:numericalresults}a)\cite{Tosi_natcomm}, 
the electronic Zeeman splitting equals the tunnel coupling
($h \gamma_\text{e} B_0 = V_\text{t}$), and 
these energy scales well exceed all other energy scales, 
including the drive frequency. 
Then, similarly to the case of magnetic driving discussed above, 
the electron dynamics is adiabatic.
Assuming that the electron is in its ground state when the 
driving starts, 
the time dependence of the electron wave function $\psi(t)$ can 
be described by the instantaneous ground state of 
$H_\text{o} + H_{B,\text{e}} +  H_{\mu,\text{e}} + H_\text{d}(t)$.
We obtain an explicit expression for $\psi(t)$ using 
second-order perturbation theory in 
$H_{\mu,\text{e}} + H_\text{d}(t)$, and 
use that to express the time-dependent Knight field $\vec b(t)$
acting on the nuclear spin via the hyperfine Hamiltonian: 
\bean
\bar{H}_\text{hf}(t) = A \braket{\psi(t) | n_\text{d} \vec S | \psi(t)}
\cdot \vec I  \equiv
\vec b(t) \cdot \vec I. 
\eean

After dropping high-harmonic terms with 
frequencies above $f$, 
we find
$\vec b(t) = \vec b_0  + \vec b_{ac} \sin (2\pi f t)$. 
Keeping only the terms up to second (third) order in the 
small energy scales $A$, $h \gamma_\text{e} \beta d$,
$U_\text{ac}$ for $\vec b_0$ ($\vec b_\text{ac})$, 
we obtain
\bean
\label{eq:fields}
\vec b_0 =
\left(\bna{c}
    \frac{A (h \gamma_\text{e} \beta  d)}{16 V_\text{t}} \\
    0 \\
    - \frac{A}{4}
\eda\right),
\,\,\,
\vec b_\text{ac} = 
\left(\bna{c}
	\frac{A (h \gamma_\text{e} \beta  d) U_\text{ac}}{8 V_\text{t}^2} \\
	0 \\
	-\frac{A U_\text{ac}}{4 V_\text{t}}
\eda\right),
\eean
from which 
we express the third-order result for the
Rabi frequency on resonance as
\bean
\label{eq:rabi}
h f_\text{R} = \frac 1 2 
\left| 
	\frac{\vec b_0}{b_0} \times \vec b_\text{ac}
\right| =
\frac{A (h \gamma_\text{e} \beta d) U_\text{ac}}{32 V_\text{t}^2}.
\eean
This is the central result of this work, demonstrating that 
electric driving ($U_\text{ac}$) can indeed induce 
nuclear-spin Rabi oscillations
when assisted by hyperfine interaction ($A$) 
and the inhomogeneous magnetic field 
($\beta$).

The nuclear-spin Larmor frequency at the tipping point
is
$f_\text{L} \approx A/(4h) \approx 30\, \text{MHz}$, and
from experimentally feasible parameters
($d = 15\, \text{nm}$, 
$U_\text{ac}/h = 100\, \text{MHz} $ corresponding to 
$E_\text{ac} = 27.6 \, \text{V}/\text{m}$), 
we estimate a nuclear-spin 
Rabi frequency $f_\text{R} \approx 72\, \text{kHz}$ 
upon resonant driving. 
This Rabi frequency corresponds to an effective coupling constant
$f_\text{R}/E_\text{ac} \approx 
3\times 10^6 \, \text{kHz} \, \mu\text{m}/\text{V}$, 
which is more than 4 orders of magnitude larger than
the electric
 coupling determined experimentally for the P ensemble in bulk Si,
attributed to electronic g-tensor modulation\cite{Sigillito}\footnote{Further comparisons: our coupling strength estimate is 
approximately 3 orders of magnitude larger than that determined
experimentally  for the single-molecule magnet in Ref.~\cite{Thiele},
and a factor of 50 smaller than that theoretically estimated for the
combined magnetic-electric two-photon Raman scheme 
in Ref.~\cite{Tosi-nuclearspin}}.
Key ingredients in this giant enhancement are:
(i) the energy denominator which has to be bridged by the perturbation 
mechanisms is much lower in the dot-donor hybrid
(it is $h V_\text{t}\approx 4.1 \, \mu\text{eV}$ 
in our example) than for a P donor
in bulk ($\sim 10\, \text{meV}$),
(ii) the spatial extension of the electronic orbitals is much larger
 in the dot-donor system ($d = 15\, \text{nm}$) than in 
an isolated donor ($\sim 1 \, \text{nm}$).

To demonstrate the reliability of the above quantitative considerations,
we numerically solved the $8\times 8$ time-dependent Schr\"odinger
equation for the model defined in Eq.~\eqref{eq:model}.
The initial state was the ground state of the nuclear-spin 
qubit, i.e., the ground state of the static
Hamiltonian $H-H_\text{d}(t)$. 
In Fig.~\ref{fig:numericalresults}c,
we show a `Chevron' plot, i.e., the time-dependent 
occupation probability $P_2$ 
of the first excited state of the static Hamiltonian
(the excited qubit basis state).
The plot reveals regular Rabi oscillations of the nuclear spin. 
The horizontal line shows the numerically obtained 
nuclear Larmor frequency, i.e., the difference between 
the second and first eigenvalues of the static Hamiltonian. 
The vertical line shows the spin-flip time $1/(2f_\text{R})$,
evaluated from Eq.~\eqref{eq:rabi}, matching well 
with the probability peak of the numerical data. 

Electrical potential fluctuations are expected to affect
these Rabi oscillations, and thereby hinder their experimental
observability, and their application as 
single-qubit gates. 
The effect of noise is illustrated in Fig.~\ref{fig:numericalresults}d, 
where damped Rabi oscillations are shown 
for different strengths of the noise. 
Noise is modelled as a random static on-site energy
difference between the interface and the donor 
sites\cite{Tosi_natcomm}, 
$H_\text{n} = \frac{U_\text{n}}{2} \sigma_z$, 
with a Gaussian probability distribution 
$\rho_\sigma(U_\text{n})$ characterized by its 
standard deviation $\sigma$. 
Each curve $\bar{P}_2(t)$ 
in Fig.~\ref{fig:numericalresults}d
is derived by computing the coherent time evolution 
of the excited-state occupation probability $P_2(t,U_\text{n})$, 
and averaging via
\bean
\label{eq:noiseaverage}
\bar{P}_2(t) = \int_{-\infty}^{\infty}
d U_\text{n} P_2(t,U_\text{n}) \rho_\sigma(U_\text{n}).
\eean
We evaluated this integral 
numerically, using an equidistant
grid with 101 points 
in the $[-2.5,2.5]\sigma$ interval. 

The key observation in Fig.~\ref{fig:numericalresults}d
is that already the first Rabi-oscillation peak 
is strongly damped at a noise
level $\sigma/h = 3\, \text{MHz} $ (dotted).
Based on recent experiments, we expect that the 
noise level in Si devices\cite{Freeman,Tosi_natcomm} is of the order of 
$\sigma_\text{exp}/h = 0.2$ GHz,
which suggests that the observation
of these Rabi oscillations requires a significant
reduction of charge noise.

This strong noise susceptibility has a simple interpretation,
offering a vastly improved 
alternative setup.
At the tipping point, charge noise is effective
in displacing the electron, 
therefore (i) changes the electron's overlap with the P nucleus,  hence 
(ii) changes the effective strength of the hyperfine
interaction, and therefore
(iii) changes the z-directional Knight field, leading to 
(iv) a change in the nuclear-spin Larmor frequency
\footnote{The electrically induced change in the 
nuclear-spin Larmor frequency is responsible for a further
effect: an anomalous Bloch-Siegert shift\cite{BlochSiegert,Shirley,Romhanyi,suppmat}}. 
This is apparent in Eq.~\eqref{eq:fields}:
the detuning-induced shift of the Larmor frequency
is $\delta f_\text{L} = -AU_\text{n}/(4V_\text{t})$.
Since noise shifts the Larmor frequency, the electric drive
becomes off-resonant for many of the noise realizations, 
hence the noise-averaged Rabi oscillations become incomplete.
A solution to this problem, leading to orders of magnitude improvement
of the quality of Rabi oscillations, is based on using two 
electrons in the dot-donor system instead of one. 
We show here that in such a two-electron (2e) setup, 
the modulation of the transverse Knight field via
electric driving remains significant, 
whereas the modulation of the longitudinal Knight field
can be suppressed compared to the 1e case,
providing efficient control of the nuclear-spin qubit 
together with resilience to charge noise. 

For the demonstration, we 
assume that  (see, e.g., \cite{Urdampilleta}),
two electrons occupy the dot-donor system, 
and the on-site energy difference is tuned to
the tipping point between the (1,1) and (0,2)  
configurations, where $(N_\text{i},N_\text{d})$ denotes
when $N_\text{i}$ 
($N_\text{d}$) electrons reside at the interface (donor). 
The Hamiltonian is a natural 2e generalization 
of the 1e Hamiltonian given above\cite{suppmat}, 
and is represented
by a 10$\times$10 matrix due to the 5 (2) different
electron (nuclear) spin states
$\ket{S}$, $\ket{T_+}$, $\ket{T_0}$, $\ket{T_-}$, $\ket{S_{02}}$
($\ket{\Uparrow}$, $\ket{\Downarrow}$).
Fig.~\ref{fig:numericalresults}b 
shows the dependence of the energy spectrum 
on the  parameter $\tilde{U}$ representing the 
on-site energy detuning from the (1,1)-(0,2) tipping point,
using values
$V_\text{t}/h = 50 \, \text{GHz}$ and
$B = 906.5\, \text{mT}$.
We choose the nuclear-spin qubit as the two energy eigenstates 
associated
to the lowest-energy line labelled $\ket{S_\text{g}}$.
Figure \ref{fig:numericalresults}e shows that the 2e
setup is more than three orders of magnitude more
resilient to charge noise then the 1e setup. 
Our data (solid) predicts that multiple high-quality Rabi oscillations 
can be observed for the estimated experimental noise level
$\sigma_\text{exp}/h = 0.2$ GHz.
(See \cite{suppmat} for further details.)

\emph{Generalizations.}
In an actual experiment, the inhomogeneous magnetic field can 
be provided by a micromagnet\cite{PioroLadriere}. 
However, incorporating the magnet in the setup complicates
fabrication. 
We argue that, 
similarly to the case of 
electrically driven electron spin resonance\cite{Nowack,Golovach} 
it is possible to avoid that complication if
spin-orbit interaction is significant in the sample,
as highlighted for the Si conduction band in recent 
experiments\cite{Veldhorst_spinorbit,Corna,Jock}.
In a simple two-site model\cite{Danon,Boross} of the dot-donor 
setup, spin-orbit interaction
enters the electronic Hamiltonian as a spin-dependent 
tunnel coupling, e.g., 
$H_\text{so} = V_\text{s} \sigma_y S_y$.
Note that it is important that the spin projection 
appearing here is not parallel to the external magnetic field. 
It is straightforward to see that $H_\text{so}$ plays a role
analogous to the inhomogeneous magnetic field, and hence
able to mediate interaction between the driving electric field 
and the nuclear spin.
We demonstrate this for the tipping point $U=0$ in 
the 1e setup, and
weak spin-orbit interaction $V_\text{s} \ll V_\text{t}$.
Using the unitary transformation 
$W = e^{i \phi \sigma_z S_y}$
on the 
electronic degrees of freedom with
$\phi = \arctan(V_\text{s}/V_\text{t})$,
the spin-dependent tunneling term can be eliminated, 
and the same transformation renders the magnetic field
inhomogeneous. 
The leading-order terms in the transformed Hamiltonian
read $W (H_\text{o} + H_\text{so}) W^\dag \approx 
V_\text{t} \sigma_x /2$, 
and 
$W H_{B,\text{e}} W^\dag \approx 
h \gamma_\text{e} B_0 (S_z - \frac{V_\text{s}}{V_\text{t}}\sigma_z S_x)
$. 
Comparing this result with Eq.~\eqref{eq:einhom}
reveals that the spin-orbit interaction is equivalent to a
magnetic field gradient $2 \frac{B_0}{d} \frac{V_\text{s}}{V_\text{t}}$.
This implies a spin-orbit-mediated nuclear Rabi frequency 
of $f_\text{R} = \frac{A V_\text{s} U_\text{ac}}{16 h V_\text{t}^2}
\approx 72 \, \text{kHz}$
for the parameters listed above, using a spin-orbit interaction strength $V_\text{s} = 0.1 V_\text{t}$.
Another mechanism via which spin-orbit interaction influences
electron dynamics is 
g-factor anisotropy\cite{Rahman,Tosi_natcomm,Boross}, which
could also facilitate nuclear spin dynamics.

The interaction mechanisms we propose here 
between the nuclear-spin qubit and electric fields
offers potential pathways toward scalable quantum 
information processing:
it allows dispersive qubit readout using electromagnetic resonators, 
and two-qubit logical operations\cite{Tosi_natcomm} either via dipole-dipole interaction
of the donor-bound electrons, or via photon-mediated
interaction through an electromagnetic resonator. 
Note that the latter, photon-mediated interaction 
usually requires that the Larmor
frequency $f_\text{L}$
of the qubit and the mode frequency $f_\text{res}$ of the resonator
are close to each other, 
whereas the nuclear Larmor frequency in our setup (few tens of MHz) 
is way below the typical mode frequency of high-quality 
superconducting resonators (few GHz).
This difficulty could be resolved using two-photon Raman processes,
where an auxiliary classical driving field with frequency 
$\approx f_\text{res} - f_\text{L}$  compensates the
mismatch between the qubit and resonator energy quanta.
Such an arrangement has been analyzed in \cite{Tosi-nuclearspin}, 
where the classical driving field was magnetic; 
that could be substituted by all-electric driving in 
the presence of an inhomogeneous magnetic field 
or spin-orbit interaction.

\emph{Conclusions.}
In conclusion, we propose an efficient scheme to control
 dopant nuclear spins in engineered semiconductor
nanostructures with ac electric fields.
The scheme relies on an interplay of (intrinsic or artificial)
spin-orbit coupling, hyperfine interaction, and the molecular
states in a dot-donor system, and provides
orders of magnitude enhancement of the coupling strength between
the nuclear spin and the electric field, as compared to the
measured value in bulk.
We predict that even in the presence of realistic charge noise, 
high-quality nuclear-spin Rabi oscillations can be detected
in an electrically driven two-electron dot-donor system.
The  mechanism we propose could serve as an important
building block in scalable
nuclear-spin-based quantum information processing architectures.

\emph{Acknowledgments.}
We thank W. A. Coish, T. Feh\'er, A. Morello, A. Sigillito, 
G. Tosi and L. Vandersypen
for useful discussions. 
This research was supported by the National Research Development and Innovation Office of Hungary within the Quantum Technology National Excellence Program  (Project No. 2017-1.2.1-NKP-2017-00001), 
and Grants
105149,
124723,
and
108676.
B.~P.~and A.~P.~
were supported by the New National Excellence Program
of the Ministry of Human Capacities.

\bibliography{nuclearspinresonance}

\clearpage

\setcounter{equation}{0}
\renewcommand{\theequation}{S\arabic{equation}}
\setcounter{figure}{0}
\renewcommand{\thefigure}{S\arabic{figure}}
\setcounter{table}{0}
\setcounter{page}{1}
\makeatletter

\renewcommand{\bibnumfmt}[1]{[S#1]}
\renewcommand{\citenumfont}[1]{S#1}
\onecolumngrid
\parbox[c]{\textwidth}{\protect \centering \large \bf
Supplementary material for}
\begin{center}
{\bf \large ``Hyperfine-assisted fast electric control of
dopant nuclear spins in semiconductors"}
\\
\vspace{3mm}
P\'eter Boross$^1$, G\'abor Sz\'echenyi$^1$, and
Andr\'as P\'alyi$^{2,3}$
\\
\vspace{2mm}
{\it \small $^1$ Institute of Physics, E\"otv\"os University, 
1518 Budapest, 
Hungary\\
$^2$ Department of Physics, Budapest University of 
Technology and Economics, 1111 Budapest, Hungary\\
$^3$ MTA-BME Condensed Matter Research Group, 
Budapest University of Technology and Economics, 
1111 Budapest, Hungary}
\end{center}
\vspace{1cm}
\twocolumngrid


\section*{Supplementary Note 1:\\Model and results for the two-electron setup}

Here, we provide further details of the two-electron (2e) setup
discussed in the main text. 
Our goal is to express the Knight field that acts on the 
nuclear spin due to the presence of the two electrons,
and from that to deduce the parameter dependence 
of the nuclear-spin 
Rabi frequency upon resonant electric driving, in a similar fashion
as done in the main text 
for the single-electron (1e) setup, see 
Eqs.~(4), (5), (6). 

The 2e Hamiltonian is a straightforward generalization of the
1e Hamiltonian introduced in Eq.~(2) of the main text. 
Due to Coulomb repulsion between the electrons, we introduce an
extra 2e on-site Coulomb term:
\bean
H_\text{C} = \frac{U_\text{C}}{2} \left[
	n_\text{i}(n_\text{i}-1) +
	n_\text{d}(n_\text{d}-1)
	\right],
\eean
where $U_\text{C}$ is its the Coulomb energy, and we choose
to have the same Coulomb energy on both the 
$\ket{\text{i}}$ and $\ket{\text{d}}$ orbitals for simplicity. 
We keep a single orbital per site, resulting in a 6-dimensional
Hilbert space. 
We  use the 
standard singlet-triplet basis, with states denoted as 
$\ket{S(1,1)}\equiv\ket{S}$, $\ket{T_+(1,1)}\equiv\ket{T_+}$, 
$\ket{T_0(1,1)}\equiv\ket{T_0}$, $\ket{T_-(1,1)}\equiv\ket{T_-}$, 
$\ket{S(0,2)}\equiv\ket{S_{02}}$, $\ket{S(2,0)}\equiv\ket{S_{20}}$.
Here, $(N_\text{i},N_\text{d})$ denotes the charge configuration where
$N_\text{i}$ ($N_\text{d}$) electrons reside at the interface 
(donor). 

We focus on the case where the electrons are tuned by the dc
gate voltage to the vicinity of the (1,1)-(0,2) tipping point. 
There, the state $\ket{S_{20}}$ 
can be neglected due to the large Coulomb
energy $U_\text{C}$.
Then, the terms of the complete (electronic+nuclear) 
Hamiltonian of the 1e setup in Eq.~(2) of the main text
are expressed for the 2e setup in 
the product basis 
$\{\ket{S},\ket{T_+},\ket{T_0},\ket{T_-},\ket{S_{20}}\}\otimes\{\ket{\Downarrow},\ket{\Uparrow}\}$ as
\begin{subequations}
\begin{align}
	H_\text{o} &= -\tilde{U} \ket{S_{02}} \bra{S_{02}} + \frac{V_\text{t}}{\sqrt{2}} \left( \ket{S} \bra{S_{02}} + \text{h.c.} \right), \label{eq:orbitHam2e}\\
	H_{B,\text{e}} &= h \gamma_\text{e} B_0 \left(\ket{T_+}\bra{T_+}-\ket{T_-}\bra{T_-}\right), \\
	H_{\mu,\text{e}} &= 
		\frac{1}{\sqrt{2}} 
		h \gamma_\text{e} \frac{\beta d}{2} 
		\left(
			\ket{T_-}\bra{S}-\ket{T_+}\bra{S} + \text{h.c.}
		\right), 
	\\
	H_\text{hf} &= 
		\frac{A}{2}\left(-\ket{T_0}\bra{S}-\ket{S}\bra{T_0}\right. 
		\nonumber
		\\
		&\left.-\ket{T_-}\bra{T_-}+\ket{T_+}\bra{T_+}\right)I_z
		\nonumber
		\\&+
		\frac{A}{\sqrt{2}}
		\left[\left(\ket{T_+}\bra{S}+\ket{T_+}\bra{T_0}\right.\right.
		\nonumber\\
		&\left.\left.+\ket{T_0}\bra{T_-}-\ket{S}\bra{T_-}\right)I_+ + \text{h.c.}\right],\\
	H_\text{d}(t) &= -U_\text{ac} \ket{S_{02}}\bra{S_{02}}
	\sin\left(2\pi f t\right),
\end{align}
\end{subequations}
Note that here $H_\text{o}$ incorporates the Coulomb
repulsion $H_\text{C}$ as well, and 
we have defined the detuning parameter
$\tilde{U}=U-U_\text{C}$, which 
characterizes the on-site energy difference 
measured from the $(1,1)$-$(0,2)$ tipping point.

To obtain the time-dependent Knight field $\vec b(t)$ felt by the donor
nuclear spin in the 2e setup, 
we use the same adiabatic approximation that
lead us to Eqs.~(4) and (5) of the main text in the 1e setup.
For the 2e setup, we find 
\bean
\label{eq:fields2e}
\vec b_0 =
\left(\bna{c}
    \frac{A (h \gamma_\text{e} \beta  d)}{8 \delta} \\
    0 \\
    h\gamma_\text{n}B_0+\frac{A^2}{4\delta}
\eda\right),
\,\,\,
\vec b_\text{ac} = 
\left(\bna{c}
	-\frac{A (h \gamma_\text{e} \beta  d) U_\text{ac}}{16 \delta^2} \\
	0 \\
	-\frac{A^2 U_\text{ac}}{32 \delta^2}
\eda\right),
\eean
where $\delta = \frac{V_\text{t}}{\sqrt{2}}-h\gamma_\text{e}B_0$ is the detuning of $\ket{S_\text{g}}$ and $\ket{T_-}$ and we assume that $\delta \ll V_\text{t}$. Here, $\ket{S_\text{g}}$ ($\ket{S_\text{e}}$) is the ground (excited) state of the two-electron orbital Hamiltonian \eqref{eq:orbitHam2e}, see the energy spectrum 
in Fig.~2b in the main text. 
The Rabi frequency can be written as
\bean
\label{eq:rabi2e}
h f_\text{R} =
\frac{A (h \gamma_\text{e} \beta d) U_\text{ac}}{32 \delta^2}
\eean
in the $\delta\ll V_\text{t}$ limit. 
This result is almost identical to the Rabi frequency in 
the 1e case, Eq.~(6) of the main text, with the only 
difference that here $\delta$ plays the role of $V_\text{t}$. 
The highly improved noise resilience of the 2e setup
demonstrated by Fig.~2e
is due to the fact that longitudinal component of the 
ac Knight field 
[that is, the third component of $\vec{b}_\text{ac}$
in Eq.~\eqref{eq:fields2e}], 
is of third order in the small energy scales, 
in contrast to the 1e setup, where the longitudinal ac Knight field is of
second order.

The dc and ac Knight fields in \eqref{eq:fields2e}
are derived using perturbation theory for a 10-level system.
The key point in the result is the existence of a transverse Knight
field (the first component of $\vec{b}_0$),
which gains time dependence due to the ac electric 
excitation, resulting in a finite first component of $\vec{b}_\text{ac}$, 
and thereby allows for nuclear-spin control via electrical driving.
Here we show a minimal model which contains these essential 
features.
The key observation, which is somewhat counterintuitive, 
is the following: a 2e singlet state $\ket{S}$, weakly mixed with a
longitudinally polarized triplet $\ket{T_-}$, 
produces a dominantly transverse Knight field. 
To see this, consider 
$\ket{\psi}=\sqrt{1-\varepsilon}\ket{S}+\sqrt{\varepsilon}\ket{T_-}$, 
where $\varepsilon\ll1$. 
The Knight field on the donor, felt by the donor nuclear spin, is 
\begin{equation}
	\braket{\psi| n_\text{d} \vec{S}|\psi}=-(\sqrt{\varepsilon/2},0,\varepsilon/2).
\end{equation} 
In other words, even though we mix a state producing no Knight field
($\ket{S}$) and a state producing purely longitudinal Knight field 
($\ket{T_-}$), the mixture produces a dominantly transversal Knight
field. 
Importantly, this observation is harnessed in the 2e setup discussed
in the main text. 
The reason for choosing the particular parameter set of Fig.~2b and e
is to create a situation 
where the $\ket{S_g}$ and $\ket{T_-}$ are close to each other in
energy but
far from the other electronic states, and thereby 
allowing $\ket{S_g}$
to perturbatively hybridize with $\ket{T_-}$ due to 
the mixing effect of the inhomogeneous magnetic field.
This perturbative hybridization then leads to 
the transverse Knight-field components proportional 
to the magnetic-field gradient $ \beta$ 
in Eq.~\eqref{eq:fields2e}.


\section{Supplementary Note 2:\\
Bloch-Siegert shift of the resonance frequency}

A key difference between the 1e and 2e setups is the
strength of the longitudinal ac Knight field:
in the 1e setup, it arises as a strong, second-order term,
see Eq.~(5),
whereas in the 2e setup, it is a weaker, third-order term, 
see Eq.~\eqref{eq:fields2e}.
One dramatic consequence of this difference is that
the nuclear-spin Rabi oscillations of the 1e setup are much less 
resilient to charge noise, as shown in Fig.~2d,e, and discussed
in the main text.
Another observable consequence of this difference is the 
appearence of an unexpectedly large drive-strength-dependent
shift of the resonance frequency in the 1e case, which 
we refer to as an \emph{anomalous Bloch-Siegert shift}. 
As we show below, this Bloch-Siegert shift of the resonance frequency
becomes pronounced as the working point is detuned from 
the tipping point between the interface and the donor.

\begin{figure}
\centering
\includegraphics[width=1.0\columnwidth]{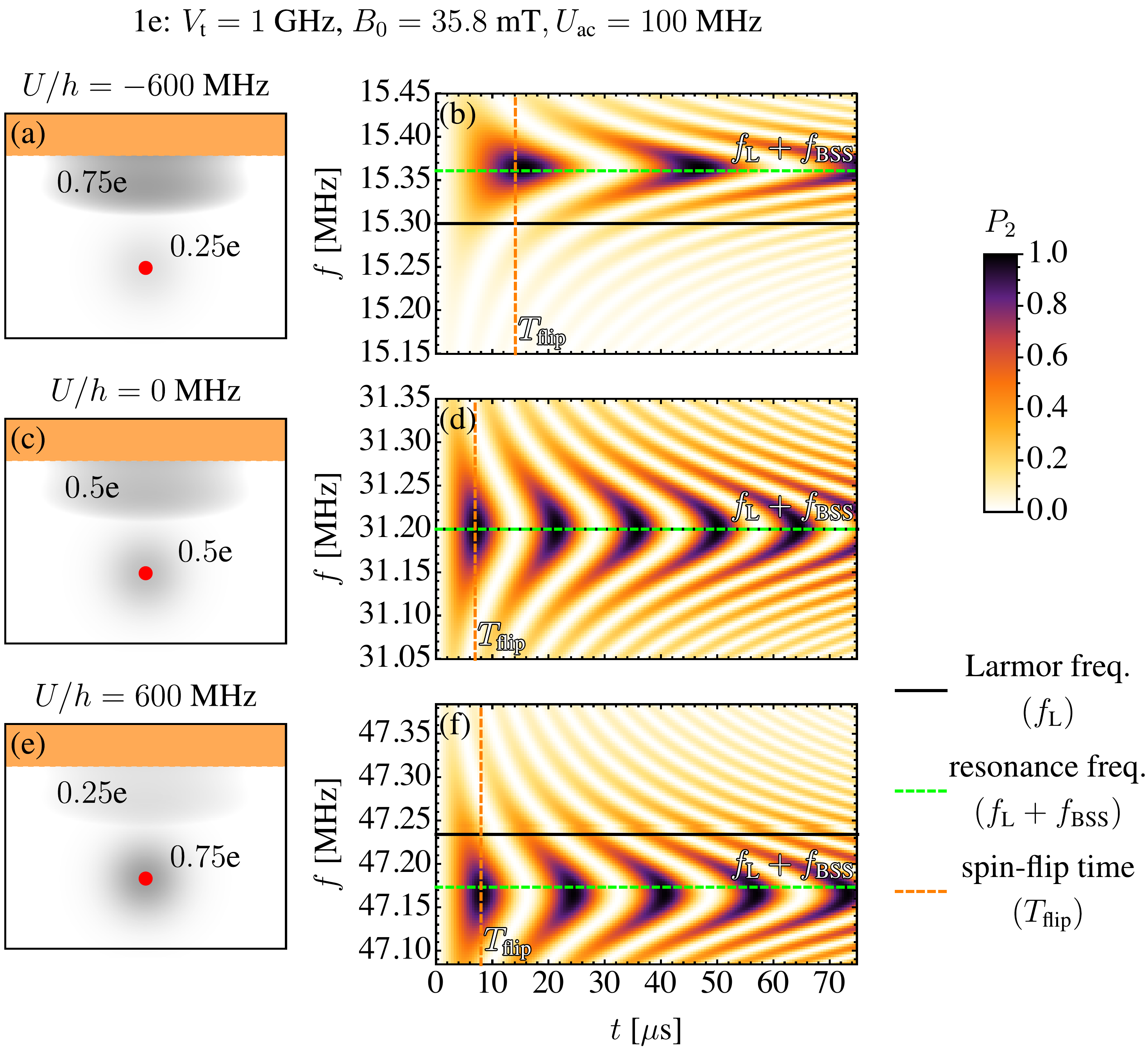}
\caption{
{\bf Anomalous Bloch-Siegert shift
in electrically driven nuclear spin resonance with a single electron.}
(a,c,e) Charge distribution of the electron in the 
dot-donor system, for a single electron, for three different values of the 
gate-induced dc electric field $U/(e d)$.
(b,d,f)
Nuclear-spin Rabi oscillations: 
occupation probability $P_2(t,f)$ 
of the excited basis state of the nuclear-spin qubit
as a function of time $t$ and drive frequency $f$,
for the charge distribution shown in
a/c/e, respectively. 
Solid black lines show the numerically calculated 
nuclear-spin Larmor frequency $f_\text{L}$, corresponding to the  
seperation between the two lowest-energy eigenvalues
of the static Hamiltonian. 
Green dashed lines show the resonance frequency
obtained as the sum of $f_\text{L}$ and the Bloch-Siegert
shift $f_\text{BSS}$ given by Eq.~\eqref{eq:bss}.
\label{fig:anomalousblochsiegertshift}
Orange vertical lines show the spin-flip time 
$1/(2f_\text{R})$ predicted by Eq.~\eqref{eq:rabigeneral}.
}
\end{figure}

Figure 
\ref{fig:anomalousblochsiegertshift}b,d,f 
show the qubit-flip probability (color coded)
in the 1e setup,
for three different dc gate voltages, whose equilibrium 
charge distributions 
are depicted in Fig.~\ref{fig:anomalousblochsiegertshift}a,c,e, 
respectively. 
The black horizontal lines in 
Fig.~\ref{fig:anomalousblochsiegertshift}b,d,f 
indicate the Larmor frequency of the nuclear-spin qubit, 
corresponding to the energy splitting of the undriven qubit. 
Fig.~\ref{fig:anomalousblochsiegertshift}b and f reveal that
the apparent
resonance frequency does not match the Larmor frequency
when the electron is detuned from the ionization point.
The deviation increases with increasing drive strength (not shown).
This effect is known as the Bloch-Siegert shift (BSS) in 
magnetic resonance [41,42], 
and was also analyzed in
electrically driven spin resonance [43].
In magnetic resonance, as long as the driving field is weaker than
the static field, and as long as the fundamental resonance
($f\approx f_\text{L}$) is considered, 
the BSS is always positive and much smaller than
the power broadening of the resonance [41,42]
(i.e., the broadening of the resonance along the driving-frequency direction),
and therefore hardly resolvable.
In our case, however, the BSS is positive in 
Fig.~\ref{fig:anomalousblochsiegertshift}b but negative in 
Fig.~\ref{fig:anomalousblochsiegertshift}f, 
moreover, the BSS is comparable to the power broadening.
Because of these unconventional features, we refer to this as
an anomalous BSS.

This anomalous BSS can be explained as a dynamical 
consequence of the electric modulation of the hyperfine
strength. 
The simple argument is that the electric drive modulates the
electron weight $n_\text{d}(U+U_\text{ac} \sin(2\pi f t))$ on the donor, 
which in turn modulates the nuclear Larmor frequency
via the hyperfine interaction
$H_\text{hf} = A n_\text{d} \vec{S}\cdot \vec{I}$, 
as the nuclear Larmor frequency 
has a Knight-field contribution of $A n_\text{d}/2$.
Taking the time average of this contribution after second-order
expansion in $U_\text{ac}/U$, we obtain
$h f_\text{BSS} = A U_\text{ac}^2 n_\text{d}''(U)/8$.
Noting that in the ground state, the electron weight on
the donor is $n_\text{d}(U) \approx \frac 1 2 \left(
1 + U/\sqrt{U^2+V_\text{t}^2}
\right)$, we conclude that
\bean
\label{eq:bss}
h f_\text{BSS} \approx - \frac{3}{16}
\frac{A U_\text{ac}^2 U V_\text{t}^2}{\left(U^2 + V_\text{t}^2\right)^{5/2}}.
\eean
This result is third order in the perturbative parameters
($A$, $U_\text{ac}$),
just like the Rabi frequency in Eq.~(6), which explains 
why the BSS is comparable to the power broadening 
in this setup. 
A good agreement is shown between 
this analytical result, represented as the horizontal green dashed
lines in Fig.~\ref{fig:anomalousblochsiegertshift}, and the numerical data. 
Note that this anomalously strong 
BSS is absent in the 2e case (not shown) discussed 
in Fig.~\ref{fig:numericalresults}, even when a finite
detuning $\tilde U$ is applied.

As a final remark, we note
 that it is possible to generalize our analytical 
Rabi frequency result
Eq.~(6) of the main text.
That result was obtained for the 1e setup, for the special case
when the working point is the tipping point between the 
interface and donor (denoted by the vertical gray line
at $U=0$ in Fig. 2a of the main text).
Allowing for a finite on-site energy 
detuning $U \neq 0$ from the tipping point, 
and following the same method as described in the main text, 
we obtain the generalization of Eq.~(6) as
\begin{equation}
h f_\text{R} = 
\frac{A (h \gamma_\text{e} \beta  d) 
U_\text{ac} V_\text{t} \left(U + V_\text{t}+\sqrt{U^2+V_\text{t}^2}\right)}{16 \left(U^2+V_\text{t}^2\right)\left(V_\text{t}+ \sqrt{U^2+V_\text{t}^2}\right)^2}.
\label{eq:rabigeneral}
\end{equation}
In Fig.~\ref{fig:anomalousblochsiegertshift}b,d,f, 
the vertical dashed orange lines indicate the 
spin-flip time $T_\text{flip} = 1/(2f_\text{R})$ based on 
Eq.~\eqref{eq:rabigeneral}, showing a satisfactory 
agreement with the peak positions of numerically
obtained excited-state occupation probabilities. 

In conclusion, here we have demonstrated that the
apparent resonance frequency in the single-electron setup depends
strongly on the drive strength when ac electrical driving is used.
This effect is interpreted as an anomalous Bloch-Siegert shift: 
in contrast to the Bloch-Siegert shift in conventional 
paramagnetic resonance, here the apparent detuning
of the resonance frequency can be comparable to 
the power broadening of the resonance, and can have an
unconventional, negative sign.




%

%
\clearpage

\end{document}